# Ultra-high-linearity integrated lithium niobate electro-optic modulators

Hanke Feng[1†], Ke Zhang[1†], Wenzhao Sun[1], Yangming Ren[2,3], Yiwen Zhang[1], Wenfu Zhang[2,3] & Cheng Wang[1*]

[1]*Department of Electrical Engineering & State Key Laboratory of Terahertz and Millimeter Waves, City University of Hong Kong, Kowloon, Hong Kong, China*

[2]*Institute of Optics and Precision Mechanics, Chinese Academy of Sciences, Xi'an, Xi'an 710119, China*

[3]*University of Chinese Academy of Sciences, 100049, Beijing, China*

*†These authors contributed equally to this article*

*\*cwang257@cityu.edu.hk*

**Abstract:** Integrated lithium niobate (LN) photonics is a promising platform for future chip-scale microwave photonics systems owing to its unique electro-optic properties, low optical loss and excellent scalability. A key enabler for such systems is a highly linear electro-optic modulator that could faithfully covert analog electrical signals into optical signals. In this work, we demonstrate a monolithic integrated LN modulator with an ultrahigh spurious-free dynamic range (SFDR) of 120.04 dB·Hz$^{4/5}$ at 1 GHz, using a ring-assisted Mach-Zehnder interferometer configuration. The excellent synergy between the intrinsically linear electro-optic response of LN and an optimized linearization strategy allows us to fully suppress the cubic terms of third-order intermodulation distortions (IMD3) without active feedback controls, leading to ~ 20 dB improvement over previous results in the thin-film LN platform. Our ultra-high-linearity LN modulators could become a core building block for future large-scale functional microwave photonic integrated circuits, by further integration with other high-performance components like low-loss delay lines, tunable filters and phase shifters available on the LN platform.

**Introduction**

Microwave photonics (MWP) is a powerful technology that makes use of optical devices and systems to perform microwave signal generation, manipulation and transmission tasks. Recent advances in integrated photonics have further propelled the realization of MWP functionalities in compact and cost-effective chip-scale systems, i.e. integrated MWP [1]. Many intriguing MWP applications have been realized in various platforms such as silicon (Si) [2,3], silicon nitride (SiN) [4,5], and indium phosphide (InP) [6,7], including broadband microwave signal generation [2,3], tunable true time delay [5], and programmable MWP filters [6]. At the heart of almost all MWP systems sits the electro-optic (EO) modulators, which are responsible for faithfully encoding signals from the microwave to the optical domain. During this process, modulation linearity, often quantified as the spurious-free dynamic range (SFDR) [8], is the most critical metric that ultimately determines the MWP link performance. A modulator with high SFDR could efficiently transmit analog microwave signals while introducing minimal unwanted spurious harmonics (e.g. third-order intermodulation distortions, IMD3) or noises.

The recently emerged thin-film lithium niobate ($LiNbO_3$, LN) platform is an excellent candidate for future high-performance integrated MWP systems [9]. The Pockels effect in LN is intrinsically linear and low-loss, ideally suited for realizing high-linearity EO modulators. Even today, the highest SFDRs are still achieved using traditional ion-diffused LN modulators that are bulky and difficult to integrate [10,11]. Compared with these bulk LN counterparts, the thin-film LN platform preserves the excellent material properties of LN, but at the same time provides much better light confinement, integrability and compactness due to its dramatically increased refractive index contrast [9]. More importantly, LN platform is now endowed with a full range of high-performance devices, including broadband frequency comb sources [12-15], ultrahigh-$Q$ microresonators [16,17], programmable filters [18], efficient frequency converters/shifters [19-21] as well as low-loss delay lines [22], which could potentially be integrated with high-linearity EO modulators on the same photonic chip for complex MWP functions. To date, many miniaturized and high-performance thin-film LN modulators have been demonstrated, exhibiting ultrahigh bandwidths beyond 100 GHz, CMOS-compatible drive voltages and low optical losses [23-29]. However, these modulators are mostly designed for digital communications only. The measured SFDRs of state-of-the-art integrated LN modulators are only ~ 100 dB·Hz$^{2/3}$ [24,25], which are far from what the material itself could possibly support, mainly restricted by the sinusoidal transfer function of the Mach-Zehnder interferometers (MZI). Realizing a high-linearity integrated LN modulator that could exploit the full potential of the LN material, will greatly boost the development of LN-based MWP systems.

To compensate for the nonlinear transfer function of a simple MZI, several linearization strategies via the implementation of novel device architectures have been proposed and adopted on various integrated photonic platforms. For example, using the ring-assisted MZI (RAMZI) scheme [30-35], SFDR values up to 111.3 dB·Hz$^{2/3}$ have been demonstrated in Si [33], and 117.5 dB·Hz$^{2/3}$ on an InP/Si heterogeneous platform [34], much higher than those achieved in simple MZI/ring modulators [36,37]. However, both Si and InP modulators rely on intrinsically nonlinear modulation mechanisms, i.e. free-carrier injection [38] and quantum confined Stark effect [39]. The linearization conditions of these modulators are therefore further complicated in order to compensate for the nonlinearities from both the device architecture and the material response. As a result, these RAMZI demonstrations have not been able to exploit the full promise of this linearization strategy – cubic terms of IMD3 cannot be fully suppressed and remain the dominant spurious harmonic species [33-35]. More complicated device configurations, such as dual-series [40] and dual-parallel [41] MZI structures, could be used to achieve the full cancellation of the cubic or even the fifth-order terms of IMD3. However, the high insertion loss and the requirement for complicated active feedback controls make these devices less competitive in practical applications. An ideal EO modulator for analog MWP systems shall feature a linear material EO response, a simple linearization strategy (e.g. RAMZI), low optical loss, high-power handing capability to minimize amplifier-induced system noises, as well as chip-scale compatibility with other functional photonic devices, simultaneously.

Here we demonstrate such an ultra-high-linearity modulator satisfying all the requirements above based on the integrated LN platform. The synergy between the excellent material properties of LN and an optimized RAMZI configuration allows us to demonstrate an ultrahigh measured SFDR of 120.04 dB·Hz$^{4/5}$ at 1 GHz, by fully suppressing the cubic terms of IMD3. The results are nearly 20 dB higher than our reference MZI modulator as well as previous record in the thin-film LN platform.

**Results**

Figure 1 shows the full microscope image and working principle of our RAMZI modulator, where a racetrack resonator with a free spectral range (FSR) of 50 GHz is coupled with one path of an MZI. When input analog RF signals containing multiple frequency tones ($f_1$ and $f_2$ in our experiment, as shown in Fig. 1a) are up converted into optical frequencies, the nonlinearity in the EO modulation process would induce unwanted spurious IMD3 components (i.e. $2f_1 - f_2$ and $2f_2 - f_1$) that are close by and could interfere with the signal frequencies (Fig. 1a). For a simple MZI biased at the quadrature point, a linear phase change $\Delta\varphi_{\text{mzi}}$ would be translated into a sub-linear intensity change, i.e. $\frac{1}{2}\left[1 + \cos\left(\Delta\varphi_{\text{mzi}} - \frac{\pi}{2}\right)\right]$, when the two arms combine, leading to a nonlinear transfer function (Fig. 1c). In our RAMZI, as the schematic diagrams in Fig. 1b-d illustrates, the modulation phase change $\Delta\varphi_{\text{mod}}$ is applied on the ring/racetrack resonator instead, which operates in the deep over-coupling state and is biased at the off-resonance point. By engineering the coupling coefficient $\kappa$ with respect to the intrinsic round-trip transmission coefficient $\tau$, the phase response of the ring resonator $\Delta\varphi_{\text{ring}}$ could be fine-tuned to achieve a super-linear relationship with the modulation phase $\Delta\varphi_{\text{mod}}$ (red curve in Fig. 1b) , which could precisely compensate for the sub-linearity of the MZI (Fig. 1c) to eliminate the cubic terms of IMD3 [30]. As a result, the output intensity of the RAMZI shows a much-linearized Fano line shape with respect to the modulation phase (Fig. 1d). In our particular device where the racetrack resonator features an intrinsic $Q$ factor of ~ $1.1\times10^6$, the required coupling coefficient $\kappa$ is 0.96 (see Methods) [30,31]. This strong coupling is achieved by an optimized 2×2 multimode interferometer (MMI) (see Methods). The devices are fabricated on a 500 nm thick LN-on-insulator substrate (see Methods for details). Modulation electrodes are placed along the two straight sections of the racetrack resonator, as shown in the inset of Fig. 1a. The measured insertion loss of the fabricated RAMZI modulator is 10.5 dB, including a 2.5-dB on-chip loss (2.2 dB from MMIs and 0.3 dB from propagation loss) and a total of 8-dB loss from fiber-to-chip coupling.

We show strongly linearized modulation transfer function in our fabricated RAMZI devices (Fig. 2). Across the measured spectrum, the device exhibits distinct line shapes depending on the phase difference between the two arms of the unbalanced MZI, $\Delta\varphi_{mzi}$ (purple dashed line of Fig. 2a), including Lorentzian, Fano, and electromagnetically induced transparency (EIT)-like line shapes. This is a result of the interference between the narrowband racetrack resonator (FSR ~ 0.4 nm) and the broadband phase response of the unbalanced MZI (FSR ~ 20 nm) according to the Fano-Anderson model [42]. Linearized Fano line shape could be clearly observed at the quadrature points of the MZI ($\Delta\varphi_{mzi}=\pi/2$ or $3\pi/2$), as shown in the insets of Fig. 2a, which agrees well with the theoretical prediction in Fig. 1d. We measure the EO transfer function of the RAMZI at this quadrature point by applying a 10-kHz triangular voltage signal and monitoring the output optical signals (red curve of Fig. 2b). As a side-by-side comparison, we also fabricate a reference MZI modulator on the same chip using similar waveguide dimensions and electrode design, the measured transfer function of which is plotted in the same figure (blue). The applied voltages in Fig. 2b are normalized by the corresponding half-wave voltages ($V_\pi$) of the respective devices for better comparison (definition of $V_\pi$ for RAMZI follows that defined in [31]). The transfer function of the RAMZI modulator clearly shows a broadened linear regime compared with that of the reference MZI modulator. This indicates that the optimized design parameters, in particular the coupling ratio of the MMI, have been precisely achieved in the fabricated devices. Figure 2c shows the scanning electron microscope (SEM) image of the fabricated 2×2 MMI structure, which is designed to be broadband and fabrication-tolerant (Methods). Figure 2d shows the simulated electric field intensity evolution within the MMI structure for fundamental transverse-electric (TE) mode, transferring $|\kappa|^2 \sim 92\%$ of the optical power from the bus waveguide into the racetrack resonator (and vice versa) for each roundtrip.

We demonstrate ultrahigh SFDR values up to 120.04 dB·Hz$^{4/5}$ in our RAMZI modulator using the measurement setup shown in Fig. 3a (see Methods for details). In contrast to most integrated MWP demonstrations on other platforms [24,25,34-36,40,41], our test link does not require an erbium-doped fiber amplifier (EDFA) after modulation due to the high-power handling capability of our LN devices, therefore minimizing the additional noise from EDFA when amplifying small signals. Figure 3b-c shows the measured output fundamental harmony (FH) and the IMD3 products as functions of the input RF powers for both the RAMZI modulator (red) and the reference MZI modulator (blue) at center signal frequencies of 1 GHz and 5 GHz, respectively. Notably, the IMD3 signals in our RAMZI show a slope of 5 in the log-log scale at both frequencies (red dashed lines), indicating that the cubic terms of IMD3 are strongly suppressed and now the fifth-order terms dominate. To the best of our knowledge, such strong suppression has not been achieved in RAMZI modulators in other material platforms [33-35], and is only made possible in a platform with an intrinsically linear material response and optimized RAMZI implementations. Since the leading spurious harmonics scale to the fifth power, the SFDR value should take a unit of dB·Hz$^{4/5}$ when normalized with the resolution bandwidth of the noise floor. Considering the noise floor of -133.8 dBm at 1 GHz during our measurement under a resolution bandwidth of 1 kHz (equivalent to -163.8 dBm/Hz), our RAMZI device shows an ultrahigh SFDR of 120.04 dB·Hz$^{4/5}$ when the optical power reaches the saturation level of our photodetector (PD) (Fig. 3b). The measured SFDR represents an 18-dB improvement over the reference MZI under the same measurement condition, and a similar level of improvement over previous integrated LN demonstrations [24,25]. This is also the highest SFDR achieved in any integrated photonic platform without active feedback controls [24,25,34-36,40,41]. At a frequency of 5 GHz (10% of the ring FSR), the measured SFDR drops to 114.54 dB·Hz$^{4/5}$ due to the lifetime limitation of the resonator (details to be discussed later), yet still representing a 13-dB improvement from the reference MZI modulator (Fig. 3c).

To further confirm that our RAMZI is indeed biased at the optimal operating points, we measure the FH and IMD3 signals at various detuning wavelengths within one FSR of the resonator, near two different quadrature points (i.e. $\Delta\varphi_{mzi} = \pi/2$ and $3\pi/2$ in Fig. 4a-b respectively). Here the two input RF tones are fixed at a relatively high power of 15 dBm near 1 GHz, such that the IMD3 products can be reliably measured. At off-resonance points *B* and *E*, which also correspond to the detuning conditions used in Fig. 3b-c, strong suppression of the IMD3 products could clearly be observed (red dashed lines), therefore giving rise to the highest carrier-to-distortion (FH/IMD3) ratio and in turn the highest SFDR (Fig. 4a-b). In contrast, other detuning points, e.g. *A*, *C*, *D*, or *E*, exhibit lower SFDR values due to significantly higher IMD3 powers, despite their higher modulation efficiencies and FH signals (blue dashed lines in Fig. 4a-b).

We show that the high-power handling capability of our thin-film LN platform is a key enabler for the high SFDRs demonstrated. Figure 4c shows the measured SFDRs at increasing received optical powers when the device is kept at the optimal operation point. Increasing the optical power elevates the FH and IMD3 products simultaneously, leading to higher interception points between these tones and larger SFDRs. In our system where SFDR is limited by the fifth-order terms, a 1-dB increment in optical power gives rise to a 1.6-dB increase in SFDR value. This trend extends until the optical power reaches the PD saturation level (10 dBm, dashed line in Fig. 4c), which corresponds to an optical power of 125 mW on chip, without the need for an EDFA after the modulator. Using a PD with higher saturation power and further increasing the optical power [12] could in principle lead to even higher SFDR values without changes to the device or the measurement setup.

**Discussions**

The record-high SFDR of 120.04 dB·Hz$^{4/5}$ in our integrated LN RAMZI modulators, achieved without the need for active feedback controls, results from a combination of the intrinsically linear EO response of LN, the high-power handling capability of the thin-film LN platform, as well as optimized RAMZI design and implementation. Even higher SFDR could possibly be achieved in the future by using PD with higher gain or saturation power, and/or adopting more advanced device configurations to cancel out the fifth-order terms of IMD3 [41]. Further implementing low-loss fiber-chip couplers [43-45] and increasing the optical powers could lead to long-distance radio-over-fiber links with high gain and high linearity simultaneously.

One limitation of the RAMZI scheme is that the SFDR value drops at higher RF frequencies due to the narrowband nature of the racetrack resonator, as our measurement and simulation results show in Fig. 4d. The RAMZI modulator offers improved SFDR over the reference MZI modulator at modulation frequencies up to 27% of the FSR (13.5 GHz in this case) according to the simulation results. Larger linearized RF bandwidths could be achieved using smaller ring/racetrack resonators with larger FSRs, at the cost of decreased modulation efficiencies (due to shorter electrode lengths) and lowered link gain.

Importantly, our ultra-linear RAMZI modulators are highly compatible with other high-performance photonic components available on the integrated LN platform, both in terms of waveguide/electrode designs and fabrication processes. Further integration of these linearized modulators with frequency comb sources, tunable filters, programmable switches and low-loss delay lines could lead to chip-scale MWP systems with unique functionalities that are not currently achievable in other photonic platforms, enabling advanced MWP applications like broadband RF signal generation, radio-over-fiber links and RF-photonic signal processing.

**Materials and methods**

**Design of the RAMZI**
The designed top widths of the optical waveguides and the racetrack resonator are both 1.2 µm to suppress excessive higher order optical modes. The racetrack resonator features an intrinsic $Q$ factor of $1.1 \times 10^6$ (measured using a reference racetrack resonator fabricated on the same chip), corresponding to an intrinsic round-trip transmission coefficient $\tau$ of 0.988. We calculate the required coupling coefficient of $\kappa \sim 0.96$ by performing a Taylor expansion to the transfer function of the racetrack resonator and using the super-linear cubic terms to exactly cancel out the sub-linear counterparts in the MZI transfer function. The unbalanced MZI used in our experiment features a length difference of 50 µm between the two arms, leading to a FSR of 20 nm.

**Design of the 2×2 MMI**
The 2×2 MMI is designed using full 3D finite-difference time domain (FDTD) simulations (Ansys Lumerical). The core MMI region has a size of 31.4 µm × 4.8 µm. The widths of the input waveguides are first tapered from 1.2 µm to 2.15 µm over a taper length of 15 µm before entering the MMI region to minimize scattering loss. The output waveguides are tapered in a similar but reversed manner. The measured coupling coefficient using a standalone MMI fabricated on the same chip agrees well with the designed value over a broad bandwidth from 1500 to 1630 nm, with a measured insertion loss of 0.8 dB.

**Device parameters and fabrication**
Devices are fabricated from commercially available 500 nm thick x-cut LN thin films (NANOLN). Optical waveguide and resonator patterns are first defined in hydrogen silsesquioxane (HSQ) using electron-beam lithography (EBL, 50 keV), and then transferred into the LN layer using optimized argon plasma-based reactive ion etching (RIE) [46]. The LN etch depth is ~ 250 nm, leaving a 250-nm-thick slab. After removing the remaining resist, the devices are cladded in a 1.8-µm-thick $SiO_2$ layer using plasma-enhanced chemical vapor deposition (PECVD). Gold electrodes and probe contact pads are subsequently formed by aligned photolithography, electron-beam evaporation and lift-off processes using AZ2070 photoresist. The positive and negative electrodes are spaced by a gap of 5 µm to ensure strong EO coupling while minimizing metal-induced optical losses. The final devices are cleaved for end-fire coupling.

**SFDR characterization**
For the two-tone test, two sinusoidal RF signals separated by 10 MHz in frequency are generated using two microwave sources (R&S SMU 200A, Hittite HMC-T2240) are injected into the modulation electrodes via a high-speed ground-signal (GS) probe (GGB industries). A continuous-wave pump laser (Santec TSL510) at 1594.9 nm is amplified by an L-band EDFA (Amonics) and sent to the RAMZI modulator using a lensed fiber after a polarization controller to ensure TE polarization. The output optical signal is collected using a second lensed fiber and sent to a 70-GHz PD (Finisar XPDV3120R). The FH and IMD3 components are then measured using a RF spectrum analyzer (R&S FSV7) at various input RF powers and frequencies. The measured noise floors are -163.8 dBm/Hz and -162.6 dBm/Hz at 1 GHz and 5 GHz, respectively, limited by the RF spectrum analyzer.

**Funding.**

National Natural Science Foundation of China (61922092); Research Grants Council, University Grants Committee (CityU 11204820, CityU 21208219, N_CityU113/20); Croucher Foundation (9509005); City University of Hong Kong (9610402, 9610455).

**Acknowledgments.**

We thank Zhaoxi Chen, Dr. Wing-Han Wong and Dr. Keeson Shum for their help in measurement and device fabrication.

**Conflict of interest.**

The authors declare no conflicts of interest.

**Data availability.**

Data underlying the results presented in this paper are not publicly available at this time but may be obtained from the authors upon reasonable request.

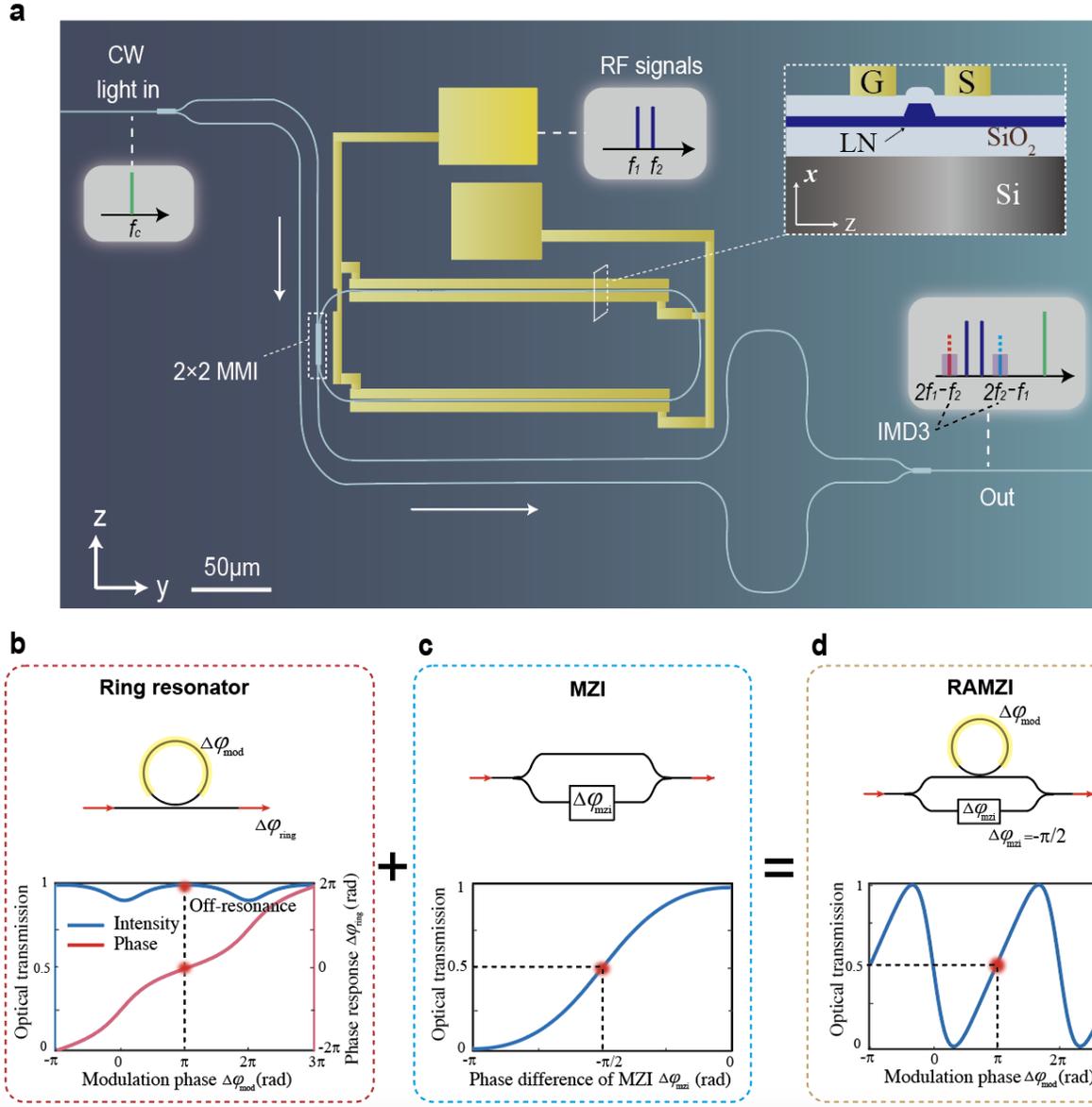

**Fig. 1. Schematic of the high-linearity RAMZI modulator. a** False-color microscope image of the fabricated LN RAMZI modulator, where a racetrack resonator is over coupled to the top path of a MZI via a 2×2 MMI coupler. Two pairs of ground-signal modulation electrodes are placed along the straight arms of the racetrack resonator to induce phase modulation. Two closely spaced RF signals ($f_1$ and $f_2$) are up converted to optical frequencies (near $f_c$) via the modulator. The RAMZI is designed to provide a linearized EO response and suppress the unwanted spurious harmonics (i.e. $2f_1$-$f_2$ and $2f_2$-$f_1$) in the output modulated optical signal. Inset shows the cross-section view of the modulation region. Scale bar: 50 µm. **b-d** Working principle of the RAMZI modulator. The ring/racetrack resonator in deep over-coupling state functions as an all-pass filter near the off-resonance point (**b**), where the output phase response $\Delta\varphi_{ring}$ exhibits a super-linear relation with modulation phase $\Delta\varphi_{mod}$ (red curve in **b**) while the intensity stays at unity level (blue curve in **b**). This super-linearity could be designed to compensate for the sub-linearity of the MZI (**c**) at the quadrature point when the two arms combine, therefore realizing a much-linearized Fano line shape in the full RAMZI device (**d**).

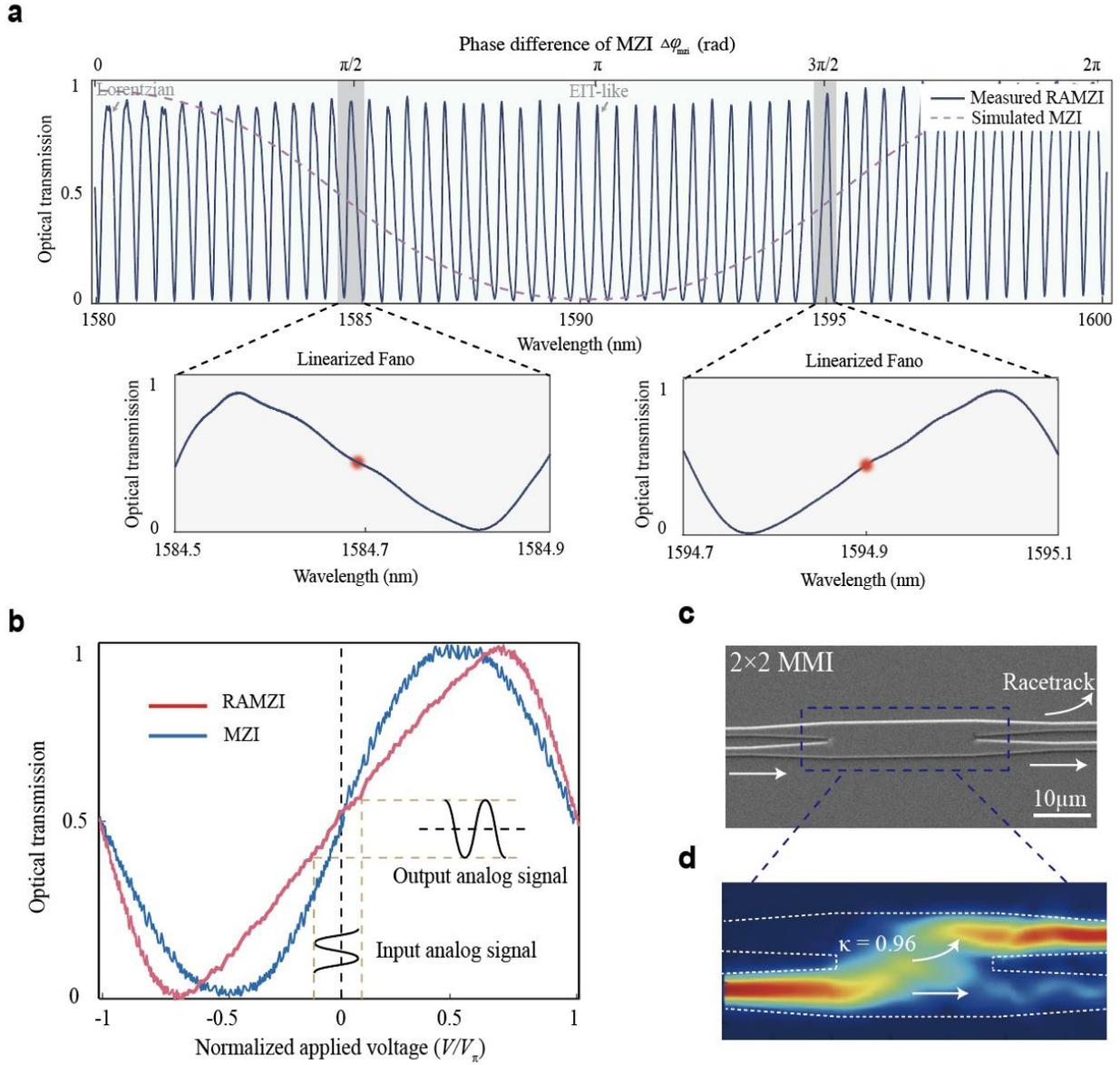

**Fig. 2. Optical characterization of the RAMZI. a** Measured optical transmission spectrum of the fabricated RAMZI device (blue solid curve), showing different line shapes depending on the phase difference $\Delta\varphi_{mzi}$ of the unbalanced MZI (purple dashed line). Linearized Fano line shape could be observed at the quadrature points of the MZI ($\Delta\varphi_{mzi}=\pi/2$ or $3\pi/2$), as shown in the insets. **b** Measured optical transmissions of the RAMZI modulator (red) and a reference MZI modulator (blue), as functions of applied voltages, showing a broadened linear regime in the RAMZI. The applied voltages are normalized by the respective half-wave voltages, $V_\pi$, for easier comparison. **c** SEM image of the 2×2 MMI structure used for achieving the deep over coupling state of the racetrack resonator. **d** Simulated electric field intensity evolution along the 2×2 MMI structure with a coupling coefficient of $\kappa = 0.96$.

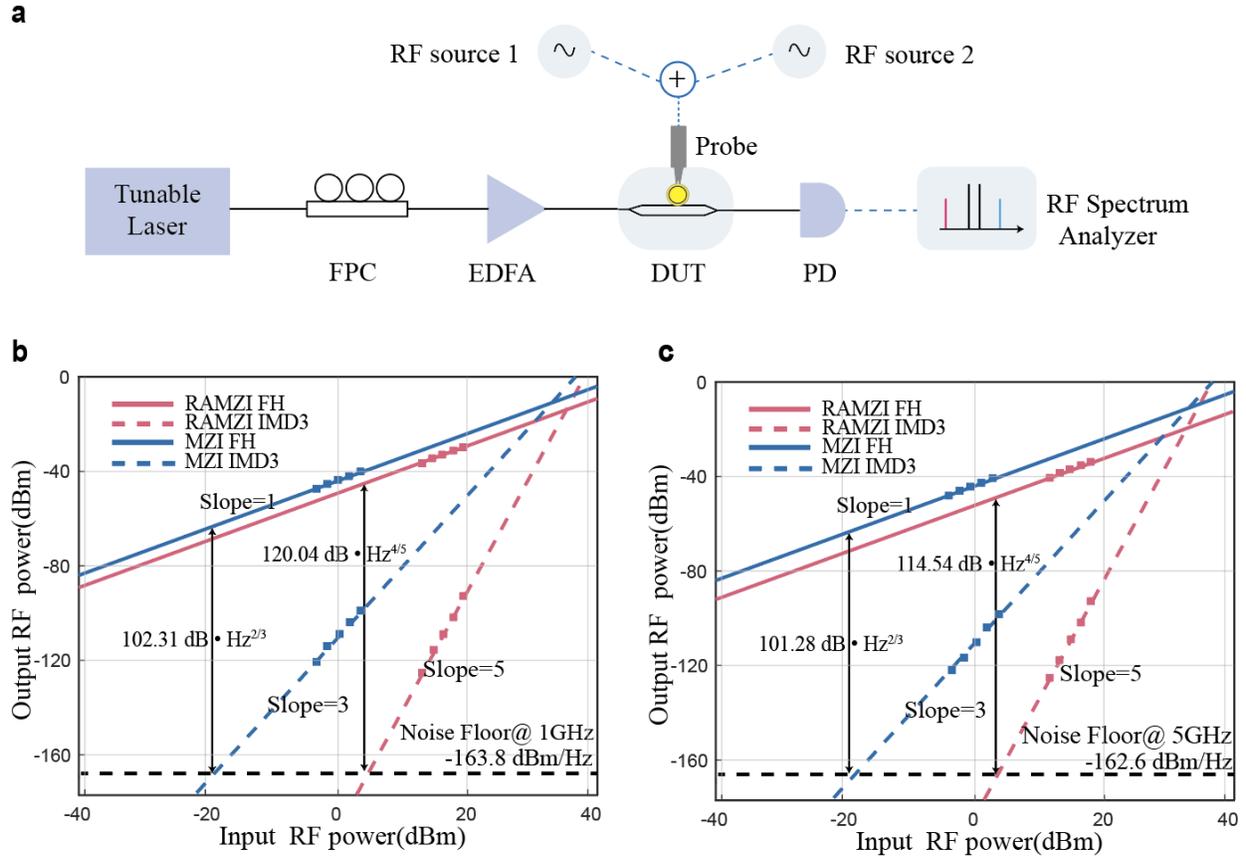

**Fig. 3. Linearity performance of the RAMZI modulator. a** Experimental setup for SFDR measurements. The test link does not require an EDFA after modulation due to the high-power handling capability of LN, therefore minimizing the extra noise from EDFA when amplifying small signals. FPC, fiber polarization controller; DUT, device under test. **b-c** Measured output FH and IMD3 powers versus input RF powers for the RAMZI (red) and the reference MZI (blue), showing high SFDR values of 120.04 dB·Hz$^{4/5}$ and 114.54 dB·Hz$^{4/5}$ at 1 GHz (**b**) and 5 GHz (**c**), respectively.

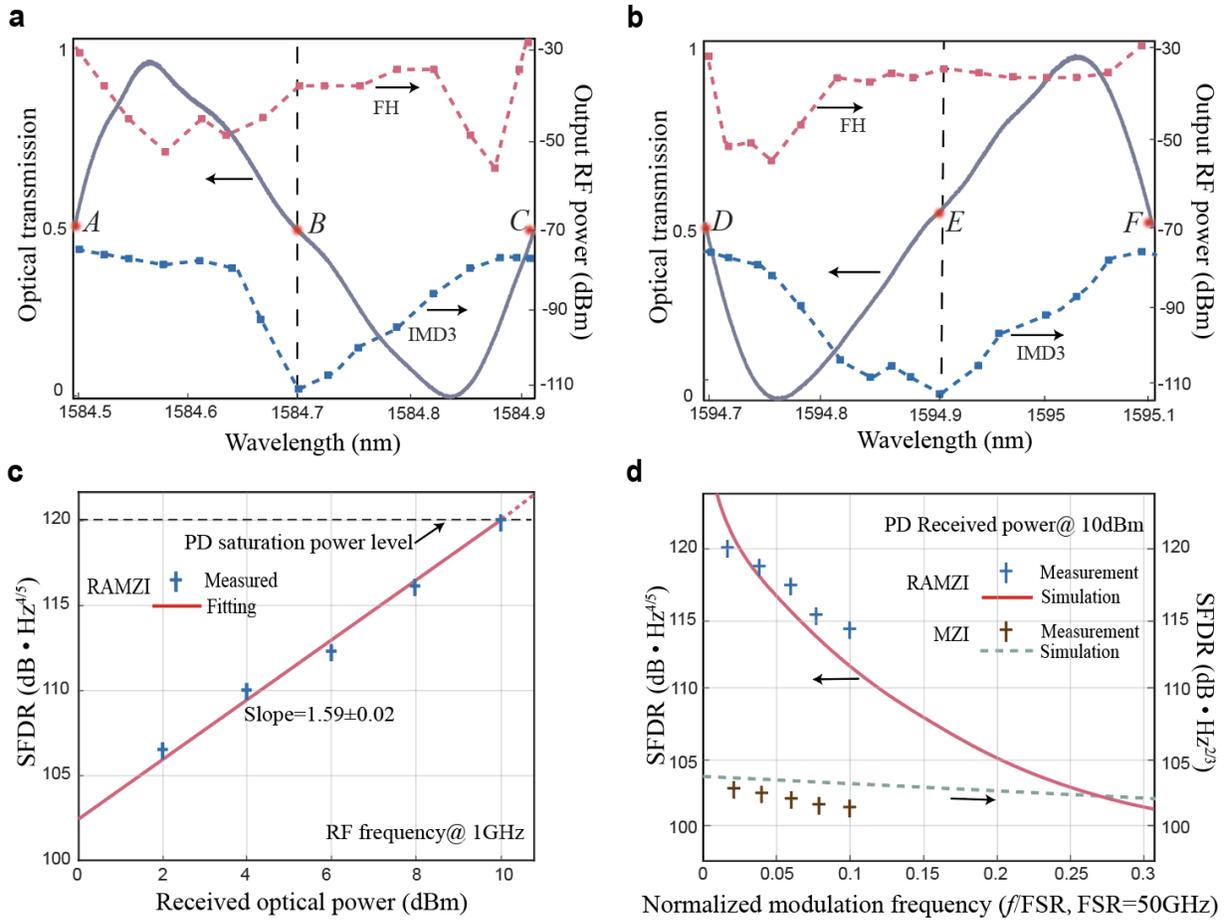

**Fig. 4. Detuning and power dependence of the RAMZI modulator**. **a-b** Measured FH (red) and IMD3 (blue) powers at various detuning wavelengths within one FSR when the MZI is biased at $\Delta\varphi_{mzi}=\pi/2$ (**a**) and $\Delta\varphi_{mzi}=3\pi/2$ (**b**), at a fixed input RF power of 15 dBm. Strong suppression of the IMD3 products could be observed at the off-resonance points (*B* and *E*). **c** Measured SFDRs at 1 GHz as a function of the received optical power, with a fitted slope of 1.59 ± 0.02 in log-log scale. The highest SFDR is measured at a received power of 10 dBm, limited by the PD saturation power (black dash line). **d** Measured and simulated SFDRs at different RF frequencies (normalized by the resonator FSR) for the RAMZI and the reference MZI.